\documentstyle[11pt,newpasp,twoside,amsmath]{article}
\begin{document}
\title{Dynamical models for dusty disk galaxies} \author{Maarten Baes
and Herwig Dejonghe} \affil{Sterrenkundig Observatorium Universiteit
Gent, Krijgslaan 281 S9, B-9000 Gent, Belgium}

\begin{abstract}
Disk galaxies contain a large amount of interstellar dust, which
affects the projection of kinematic quantities. We investigate in
detail the effects of dust extinction on the mean projected velocity
and the projected velocity dispersion. We use our results to construct
a general strategy to determine the dynamical structure of disk
galaxies, with the aim to constrain their mass distribution and
dynamical history.
\end{abstract}

\section{Introduction}

The knowledge of the dynamics of disk galaxies is essential in order
to understand their structure and history. Unfortunately, disk
galaxies are difficult systems to model dynamically, for several
reasons. One of them is the presence of a large amount of interstellar
dust, which obscures the light along the lines-of-sight. Using
extended radiative transfer models it is nowadays possible to recover
quite accurately the three-dimensional light and dust distribution in
disk galaxies (Kylafis \& Bahcall~1987, Xilouris et al.~1999). But
also the observed kinematics are affected by dust obscuration. Indeed,
each element along a line-of-sight carries its own kinematic
information, and the projected kinematics are a weighted mean of all
these contributions. We adopt the technique outlined in Baes et
al.~(2000a,b) in order to investigate in detail the effects of dust
extinction on the mean projected velocity $\bar{v}_p$ and the
projected velocity dispersion $\sigma_p$.

\section{The modelling and results}

We adopt a galaxy model which consists of a double exponential disk
and a de Vaucouleurs bulge. We construct a dynamical model (i.e.\ a
potential and a phase-space distribution function) for this galaxy. We
choose a potential that gives rise to a flat rotation curve and
represents a halo-disk structure (Batsleer \& Dejonghe 1994). Using
the Quadratic Programming modelling procedure (Dejonghe 1989) we then
construct a two-integral distribution function that is consistent with
the light density. We add a double exponential dust disk to this
model. Finally, the dust-affected $\bar{v}_p$ and $\sigma_p$ can be
calculated for various values of the inclination and optical depth.

For galaxies which are face-on or moderately inclined, the effects of
dust extinction on $\bar{v}_p$ and $\sigma_p$ are negligibly small. In
the edge-on case, the dust-affected $\bar{v}_p$-profile tends to
apparent solid body rotation, as we only see the stars moving on the
outer near edge of the disk. In meanwhile, the projected dispersion
decreases drastically as a function of optical depth for the inner
lines-of-light, as dust obscuration strongly reduces the contribution
of the high random motions of the bulge stars. Both effects are
critically dependent on inclination, and they are already much weaker
for galaxies which are only a few degrees from exactly edge-on (see
also Bosma et al.~1992).

\section{Conclusion: dynamical modelling of disk galaxies}

From our results it is clear that the effects of dust obscuration on
$\bar{v}_p$ and $\sigma_p$ are negligible for moderately inclined
galaxies. Hence it is quite safe to neglect dust extinction in the
interpretation of projected kinematics. This leads us to propose the
following strategy to construct dynamical models for disk
galaxies. Intermediately inclined disks are the best choice, as
spectra at different position angles will then show different
projections of the velocity ellipsoid.

First, one should determine the three-dimensional light distribution
of the galaxy, using deprojection techniques which take the dust into
account. The accuracy of the results can be tested by comparing models
in different wavebands with the galactic extinction curve (Xilouris et
al.~1999) or by comparing the derived extinction profile with
FIR/submm emission (Alton et al.~2000). Then, a set of potentials
which are consistent with the rotation curve and the light
distribution need to be determined. For each potential a
three-integral model can be constructed. Input for the fit should be
the light density and the projected kinematics along (at least) both
major and minor axes. The goodness of fit of the different models can
then be used to constrain the set of possible potentials, which will
reveal the mass distribution in the galaxy. The velocity field can
then be analysed, in particular the behaviour of the velocity
ellipsoid. This can shed a light on the mechanism responsible for the
dynamical history of the disk (Jenkins \& Binney~1990, Gerssen et
al.~1997, 2000).

\end{document}